\documentclass[notitlepage,12pt]{article}
\usepackage{makeidx}
\usepackage{amssymb}
\usepackage{amsfonts}
\usepackage{amsmath}
\usepackage{geometry}
\usepackage[page]{appendix}
\usepackage[doublespacing]{setspace}

\newtheorem{theorem}{Theorem}

\newtheorem{definition}{Definition}
\newtheorem{example}{Example}

\newtheorem{proposition}{Proposition}

\newenvironment{proof}[1][Proof]{\noindent\textbf{#1.} }{\ \rule{0.5em}{0.5em}}
\newcommand{\be}{\begin{equation}}
\newcommand{\ee}{\end{equation}}
\newcommand{\bes}{\begin{equation*}}
\newcommand{\ees}{\end{equation*}}
\newcommand{\bea}{\begin{eqnarray}}
\newcommand{\eea}{\end{eqnarray}}
\newcommand{\beas}{\begin{eqnarray*}}
\newcommand{\eeas}{\end{eqnarray*}}
\input{tcilatex}
\begin{document}

\title{Multi-unit Auctions with Budget Constraints\thanks{%
We thank NSF (SES 0752931) and Google Research for support. We also thank
seminar participants in Google Research and Microsoft Research for useful
comments.}}
\date{\today}
\author{Isa E. Hafalir, R. Ravi, and Amin Sayedi\thanks{%
Tepper School of Business, Carnegie Mellon University, Pittsburgh, PA 15213,
USA. E-mails: ssayedir@cmu.edu, isaemin@cmu.edu, ravi@cmu.edu}}
\maketitle

\begin{abstract}
Motivated by sponsored search auctions, we study multi-unit auctions with
budget constraints. In the mechanism we propose, Sort-Cut, understating
budgets or values is weakly dominated. Since Sort-Cut's revenue is
increasing in budgets and values, all kinds of equilibrium deviations from
true valuations turn out to be beneficial to the auctioneer. We show that
the revenue of Sort-Cut can be an order of magnitude greater than that of
the natural Market Clearing Price mechanism, and we discuss the efficiency
properties of its ex-post Nash equilibrium.

\textbf{Keywords:} Multi-Unit Auctions, Budget Constraints, Sponsored Search

\end{abstract}

\section{Introduction}

Consider the advertisement department of a computer manufacturer, who wants
to appear in a particular Web search engine's query for \textquotedblleft
laptops.\textquotedblright\ Search engines use complicated rules to
determine the allocation\footnote{%
By allocations, we mean which advertisements will be displayed, and in which
order they will be displayed if there is more than one.} of these
advertisements, or \textquotedblleft sponsored links,\textquotedblright\ and
also their pricing rules. Roughly, the advertisers specify \textquotedblleft
a value per click\textquotedblright\ and a daily maximum budget. Allocation
and pricing is then determined by a complex algorithm that makes sure that
the advertisers are not charged more, per click, than their stated values
and also are not charged more than their total budget per day.\medskip 

Advertisers' true (estimated) values per click and daily budgets are, of
course, their private own information, and given any allocation and pricing
rule they will act strategically in bidding their values and budgets. It is
then natural to ask whether there is any mechanism in which the participants
would prefer to reveal their types truthfully: in this model, their per
click values and daily budgets. Then there will not be any \textquotedblleft
gaming\textquotedblright\ of the mechanism and socially efficient
allocations can be implemented. Second-price auctions in single-unit auction
problems and different versions of Vickrey-Clark-Groves mechanisms in more
general setups have been very successful in implementing socially efficient
allocations in \textquotedblleft dominant strategies.\textquotedblright\
Unfortunately, a recent impossibility result (Dobzinski et al. 2008)
precludes the existence of a truthful mechanism with Pareto-optimal
allocations in this important setting.\medskip 

In this paper, we propose a mechanism--\emph{Sort-Cut}--which yields good
revenue and Pareto optimality properties. In our mechanism, understating
budgets or values is weakly dominated. Thus the only way a bidder can
possibly benefit from lying in our mechanism is by overstating her values or
budgets. We also show that the revenue of Sort-Cut is nondecreasing in
budgets and values, which in turn yields high revenue for the auctioneer at
equilibria.\medskip 

The idea of Sort-Cut is very similar to the idea of a second-price auction.
In second-price auctions without budget constraints, the highest bidder is
allocated the object and pays the highest loser's bid to the auctioneer.
Uniform-price auctions generalize this idea to multi-unit auctions. The idea
is to charge the winners the opportunity cost: the losers' bids. When the
bidders have budget constraints, however, losers might not be able to buy
all the items if offered: they might simply not be able to afford it. Taking
this into account, we modify the algorithm to charge the winners, per item,
the value of the highest-value loser, but only up to this loser's budget.
After the highest-value loser's budget is exhausted, she would not be able
to afford any more items, so we start charging the winners the
second-highest loser's value, up to her budget and so on. Given this pricing
idea,\footnote{%
There is a caveat here, which is that the lowest-value winner might not be
able to exhaust all of her budget. Then all higher-value bidders are charged
first at the lowest-value winner's value up to her unused budget. This makes
sense as the lowest-value winner is still a competitor with other winners to
buy further items. The pricing for the lowest-value winner, for the same
reason, starts from the highest-value loser. She cannot be a competitor with
herself.} the winners and losers are determined via a cut-point to clear the
market, i.e. all the available items are sold.\medskip 

Sort-Cut has a number of desirable properties. First of all, it sells all
the items so there is no inefficiency in that sense (whereas Borgs et al.
2005 and Goldberg et al. 2001 might leave some of the items unallocated).
Second, bidders can only benefit by overstating their values or budgets, a
deviation that is the most desirable for the auctioneer. Third, allocation
in the equilibrium of Sort-cut is nearly Pareto optimal in the sense that
all winners' values are greater than the cut-point bidder's announced value,
and all losers' values are smaller than the cut-point bidder's announced
value. And lastly, Sort-Cut reduces to a second-price auction when there are
no budget constraints.\footnote{%
Generalized second-price auctions studied by Edelman et al. (2007) also has
a similar idea in multi-item auctions. In that mechanism the winner of the
best item (first sponsored link) is charged the bid of the second-best item,
the winner of the second-best item is charged the bid of the third-best
item, and so on. In this environment there are no budget constraints and the
second-highest bid is always the competitor of the highest value.}\medskip 

We show that the revenue of Sort-cut differs at most by the budget of one
bidder from the revenue of the \textquotedblleft \emph{market clearing price
mechanism}.\textquotedblright\ The market clearing mechanism determines a
market clearing price and sells all the units for that price. This
mechanism, however, is not truthful, and the bidders can benefit from
understating their budgets (thereby decreasing the auctioneer's
revenue).\medskip 

After discussing the related literature below, we introduce the model and
our mechanism in Section 2. Section 3 discusses the truthfulness, revenue,
and Pareto optimality properties of Sort-Cut. Section 4 compares the market
clearing price mechanism with Sort-Cut. Finally, Section 5 discusses
possible extensions of the current model.\medskip 

\subparagraph{Related Literature}

The problem of multi-unit auctions with budget-constrained bidders was
initiated by Borgs et al. (2005). Our model is similar to theirs except that
we need not assume that utility is $-\infty $ when budget constraints are
violated. They introduce a truthful mechanism that extracts a constant
fraction of the optimal revenue asymptotically; however, their mechanism may
leave some units unsold. The idea is to group the people randomly into two
groups, and use the market clearing price of each group as a posted price
for the other group. Another paper that uses the same model is by Abrams
(2006), which uses techniques similar to Borgs et al. (2005) but improves
upon them; however, it may still leave some units unsold.\medskip 

In an important paper, Dobzinski et al. (2008) prove an impossibility
result. They first assume that the budgets of all players are publicly
known, and under this assumption, they present a truthful mechanism that is
individually rational and Pareto-optimal. Their mechanism essentially
applies Ausubel's multi-unit auction (Ausubel 2004) to this budgeted
setting. Then they show that their mechanism is the unique mechanism that is
both truthful and Pareto-optimal under the assumption of publicly known
budgets. Finally, by showing that their mechanism is not truthful if budgets
are private knowledge, they conclude that no mechanism for this problem can
be individual rational, truthful, and Pareto-optimal.\medskip 

Bhattacharya et al. (2010a) show that although the mechanism proposed for
public budgets in Dobzinski et al. (2008) is not truthful, for lying to be
beneficial the bidder must overstate her budget (value may be overstated or
understated). This, together with the fact that the utility of a bidder who
is charged more than her budget is $-\infty $, allows them to modify the
non-truthful deterministic mechanism into a truthful randomized mechanism.
For each bidder, instead of charging her the price specified by Dobzinski et
al. (2008), they run a lottery (with appropriate probability) and charge her
either $0$, or all of her announced budget. Therefore, since a bidder has to
pay all of her announced budget with positive probability, the expected
utility of over stating the budget becomes $-\infty $. The assumption of the
utility being $-\infty $ when the budget constraints are violated does not
seem very realistic. In our work, we drop that assumption by assuming that
the utility of bidder who has to pay more than her budget is an arbitrary
negative value. Furthermore, we avoid randomized pricing and allocation to
guarantee ex-post individual rationality.\medskip 

Ashlagi et al. (2010) look at budget constraints in position auctions; in
their setting, bidders must be matched to slots where each slot corresponds
to a certain fraction of the total supply. Bidders are profit maximizers who
face budget constraints. They assume that a violation of budget constraints
leads to zero utility for the bidder. They propose a modification of the
Generalized Second Price mechanism that is Pareto-optimal and envy free. In
their setting, the fraction of supply on each of the slots is fixed; this
makes their problem more like a matching problem with a discrete structure.
However, in our setting, the auctioneer has complete freedom regarding how
much of the supply to give to each of the bidders.\medskip 

Other papers have studied budget constraints in mechanism design, but in
settings quite different from ours. Feldman et al. (2008) give a truthful
mechanism for ad auctions with budget-constrained advertisers where there
are multiple slots available for each query, and an advertiser cannot appear
in more than one slot per query. They define advertisers to be
click-maximizers; i.e. advertisers do not value their unused budget, they
just want to maximize the amount of supply they get. However, in our model,
advertisers are profit-maximizers.\medskip 

Pai and Vohra (2010) look at optimal auctions with budget constraints. In
their setting, there is one indivisible good to be allocated, making the
setting naturally different from ours. Moreover, they assume $-\infty $
utility if budget constraints are violated. In another paper, Malakhov and
Vohra (2008) look at the divisible case; however, they assume that there are
only two bidders, one of whom has no budget constraints while the budget
constraint of the other one is publicly known. Kempe et al. (2009) look at
budget constraints when the bidders have single-unit demand and items are
heterogenous. Bhattacharya et al. (2010b) show that a sequential posted
price can achieve a constant fraction of the optimal revenue in a budgeted
setting with heterogeneous items when the budgets are common
knowledge.\medskip 

Both Borgs et al. (2005) and Dobzinski et al. (2008) argue that lack of
quasi-linearity (because of hard budget constraints) is the most important
difficulty of the problem. Still, some papers have tried to solve the
problem by relaxing the hard budget constraints (Maskin 2000), or by
modeling the budget constraint as an upper bound on the value obtained by
the bidder rather than her payment (Mehta et al. 2007). It has also been
shown (Borgs et al. 2005) that modeling budget constraints with quasi-linear
functions can lead to arbitrarily low revenue.\medskip 

Benoit and Krishna (2001) study an auction for selling two single items to
budget-constrained bidders. They focus mainly on the effect of bidding
aggressively on an unwanted item with the purpose of depleting the other
bidder's budget. A similar effect arises in our model, but the focus of our
work is generally very different from theirs. Another paper is that by Che
and Gale (1996), which compares first-price and all-pay auctions in a
budget-constrained setting and shows that the expected payoff of all-pay
auctions is better under certain assumptions. However, they do not consider
multi-unit items.

\section{The Model and Sort-Cut}

There are $m$ divisible units of a good for sale. There are $n$ bidders with
a linear demand up to their budget limits. Specifically, each bidder $i\in
N=\{1,...,n\}$ has a two-dimensional type $\left( b_{i},v_{i}\right) $ where 
$b_{i}$ denotes her budget limit and $v_{i}$ denotes her private value.
Bidder $i$'s utility by getting $q$ (possible fractional) units of the good
and paying $p$ is given by%
\begin{equation*}
u_{i}\left( q,p\right) =\left\{ 
\begin{tabular}{ll}
$qv_{i}-p$ & if $p\leq b_{i}$ \\ 
$-C$ & if $p>b_{i}$%
\end{tabular}%
\right. 
\end{equation*}%
where $\infty \geq C>0.$\medskip 

We are interested in mechanisms to sell $m$ units to $n$ bidders which have
good incentive, efficiency, and revenue properties. The equilibrium concept
we use is that of an \emph{ex-post Nash equilibrium}.\ In an ex-post Nash
equilibrium, no bidder wants to deviate after she observes all other
players' strategies. We believe that this is an appropriate equilibrium
concept, as we are motivated by sponsored search auctions. Typically,
sponsored search auctions are dynamic auctions; and bids can be changed at
any time. Therefore, it is reasonable that in a stable situation (steady
state), no bidder will want to deviate even after the bids are revealed.
Since the equilibrium concept is ex-post Nash, we do not need to assume
strong conditions on private information. Specifically, we can allow for
interdependency in two dimensional type within or across bidders.\medskip 

We focus on direct mechanisms in which bidders announce their types (values
and budgets). A mechanism consists of an allocation rule (how many units to
allocate to each bidder) and a pricing rule (how much to charge each
bidder). It takes the announcements as inputs and produces an allocation and
a pricing scheme as an output. We consider mechanisms that satisfy two
properties: (i) it must sell all $m$ units, and (ii) it must satisfy
individual rationality constraints (i.e. all bidders prefer to participate
in the mechanism). Note that the latter condition implies that bidders who
are not allocated any units (losers) cannot be charged a positive price.
Bidders who are allocated nonzero units (winners), however, will be charged
a positive price.\ Let us first introduce a general and an abstract \emph{%
pricing rule}.

\begin{definition}
The price is set according to a pricing function $\alpha :%
%TCIMACRO{\U{211d} }%
%BeginExpansion
\mathbb{R}
%EndExpansion
_{+}\rightarrow 
%TCIMACRO{\U{211d} }%
%BeginExpansion
\mathbb{R}
%EndExpansion
_{+}$, if the marginal price of the next unit is $\alpha \left( y\right) $
dollars for a buyer who has already spent $y$ dollars in the market. In
other words, if the pricing of an item is set according to $\alpha ,$ a
buyer with $b$ dollars can afford%
\begin{equation*}
x\left( \alpha ,b\right) =\int_{0}^{b}\frac{1}{\alpha \left( y\right) }dy
\end{equation*}%
units of the item. We are interested in pricing rules $\alpha \left( \cdot
\right) $ that are nonincreasing and positive.\ Hence, we assume $\alpha
\left( y\right) \leq \alpha \left( y^{\prime }\right) $ for all $y\geq
y^{\prime }$ and also $\alpha \left( y\right) >0$ for all $y.$
\end{definition}

The following definition is also convenient for later discussions.\medskip 

\begin{definition}
(Shifted pricing) For a given pricing function $\alpha $ and a positive real
number $z$, we define the pricing function $\alpha ^{z}\left( y\right) $ as: 
\begin{equation*}
\alpha ^{z}\left( y\right) =\alpha \left( z+y\right).
\end{equation*}
\end{definition}

Less formally, $\alpha ^{z}\left( y\right) $ is the pricing function
obtained by shifting $\alpha \left( y\right) ,$ $z$ units to right. Note
that we have, for any $z\in \left[ 0,b\right] ,$%
\begin{equation*}
x\left( \alpha ,b\right) =x\left( \alpha ,z\right) +x\left( \alpha
^{z},b-z\right) .
\end{equation*}

Throughout the proofs of our results, we sometimes make use of the terms
\textquotedblleft better (or worse) pricing function\textquotedblright\ and
\textquotedblleft getting to lower prices.\textquotedblright\ We say that $%
\alpha $ is a \emph{better pricing function} than $\alpha ^{\prime }$ for a
bidder if $\alpha \left( y\right) \leq \alpha ^{\prime }\left( y\right) $
for all $y.$ We say that $\alpha $ \emph{gets to lower prices} than $\alpha
^{\prime }$ for a bidder with budget $b$ if marginal payment at $b$ is lower
with $\alpha $ than with $\alpha ^{\prime }.$\medskip 

We are now ready to introduce a special class of pricing and allocation
rules, which we name \emph{Procedure Cut}.

\begin{definition}
Procedure Cut takes budgets and values of the bidders $\left( \mathbf{b},%
\mathbf{v}\right) \in 
%TCIMACRO{\U{211d} }%
%BeginExpansion
\mathbb{R}
%EndExpansion
_{++}^{n}\times 
%TCIMACRO{\U{211d} }%
%BeginExpansion
\mathbb{R}
%EndExpansion
_{++}^{n}$, a pricing rule $\alpha \left( \cdot \right) ,$ and a real number 
$c\in (0,\sum\nolimits_{i=1}^{n}b_{i}]$ as input. First, it sorts bid and
value vectors $\left( \mathbf{b},\mathbf{v}\right) $ in nonascending%
\footnote{%
It breaks ties among equal valued bidders arbitrarily.} order of values and
reindexes them so that $v_{1}\geq v_{2}\geq ...\geq v_{n}.$\footnote{%
Note that after reindexing, budgets are not necessarily sorted in a
descending way. A bidder with a high valuation could have a small budget.}
Then, it picks $j$ such that $c\leq \sum\nolimits_{i=1}^{j}b_{i}$ and $%
c>\sum\nolimits_{i=1}^{j-1}b_{i}$. Let $s=\sum\nolimits_{i=1}^{j}b_{i}-c$.
Procedure Cut sets the pricing function of bidders $1,...,j-1$ to $\alpha
^{c}$ and the pricing function of bidder $j$ to $\alpha ^{c+s}.$ The
allocation of each bidder $1,..,j-1$ is such that she spends all her budget,
i.e. $x_{i}=x\left( \alpha ^{c},b_{i}\right) $ for $i=1,..,j-1.$ The
allocation of bidder $j$ is such that she spends $b_{j}-s$ of her budget,
i.e. $x_{j}=x\left( \alpha ^{c+s},b_{j}-s\right) .$ Bidder $j$'s unused
budget is denoted by $s,$ where $s\in \lbrack 0,b_{j}).$ All bidders $%
j+1,...,n$ get no allocation and pay nothing.

Define $X\left( c,\left( \mathbf{b},\mathbf{v}\right) \right) $ to be the
total number of units allocated to all bidders, i.e. $X\left( c,\left( 
\mathbf{b},\mathbf{v}\right) \right) =\sum\nolimits_{i=1}^{j}x_{i}.$ Bidders 
$1,...,j$ are called full winners, bidder $j$ is called a partial winner (or
a cut-point bidder), and bidders $j+1,...,n$ are called losers.
\end{definition}

We consider pricing rules that are not too high, in the sense that they will
be able to sell all the items if all budgets are exhausted. Hence we assume
that for $B\equiv \sum\nolimits_{i=1}^{n}b_{i}$%
\begin{equation*}
\alpha \left( B\right) \leq \frac{B}{m}.
\end{equation*}%
With this assumption, we can easily conclude that $X\left( B,\left( \mathbf{b%
},\mathbf{v}\right) \right) \geq m.$ This is because when $c=B,$ all bidders
are full winners and their allocations satisfy 
\begin{equation*}
x\left( \alpha ^{B},b_{i}\right) \geq \frac{b_{i}}{\frac{\sum%
\nolimits_{i=1}^{n}b_{i}}{m}}
\end{equation*}%
and hence 
\begin{equation*}
X\left( B,\left( \mathbf{b},\mathbf{v}\right) \right)
=\sum\nolimits_{i=1}^{n}x\left( \alpha ^{B},b_{i}\right) \geq m
\end{equation*}

We are interested in rules that sell $m$ units. In the following
proposition, we show that for any procedure cut rule, $X\left( c,\left( 
\mathbf{b},\mathbf{v}\right) \right) $ is strictly increasing and continuous
in $c.$ Together with the assumption that $X\left(
\sum\nolimits_{i=1}^{n}b_{i},\left( \mathbf{b},\mathbf{v}\right) \right)
\geq m,$ this will imply that there will be a unique $c$ such that $X\left(
c,\left( \mathbf{b},\mathbf{v}\right) \right) =m.$

\begin{proposition}
\label{prop1}$X\left( c,\left( \mathbf{b},\mathbf{v}\right) \right) $ is
strictly increasing and continuous in $c.$
\end{proposition}

\begin{proof}
In the Appendix.\medskip 
\end{proof}

As noted above, an important corollary of Proposition 1 is that there will
be a unique $c^{\ast }$ that will satisfy $X\left( c^{\ast },\left( \mathbf{b%
},\mathbf{v}\right) \right) =m.$

\begin{definition}
We call the unique $c^{\ast }$ with $X\left( c^{\ast },\left( \mathbf{b},%
\mathbf{v}\right) \right) =m$ to be the \emph{cut-point}. Given pricing
function $\alpha \left( \cdot \right) $ and vectors $\left( \mathbf{b},%
\mathbf{v}\right) ,$ we name Procedure Cut that sells $m$ items (with $%
c=c^{\ast }$) to be the $m$\emph{-Procedure Cut}.
\end{definition}

We now can introduce our new mechanism that we call the \emph{Sort-Cut
Mechanism}.

\begin{definition}
Sort-Cut is a $m$-Procedure Cut mechanism in which $\alpha \left( \cdot
\right) $ is a step function defined by (reindexed) $\left( \mathbf{b},%
\mathbf{v}\right) $: $\alpha \left( y\right) =v_{i}$ for $y\in
(\sum\nolimits_{k=1}^{i-1}b_{k},\sum\nolimits_{k=1}^{i}b_{k}].$\footnote{%
And also $\alpha \left( y\right) =\varepsilon >0$ for $y\in (B,\infty )$}
\end{definition}

In other words, Sort-Cut takes the vectors $\left( \mathbf{b},\mathbf{v}%
\right) $ and sorts them in nonascending order of values, calculates the
unique cut-point $c^{\ast }$ according to the pricing function that each
full winner (bidders $1,..,j-1$) pays $v_{j}$ per unit up to a budget of $s,$
then pays $v_{j+1}$ per unit up to a budget of $b_{j+1},$ then pays $v_{j+2}$
per unit up to a budget of $b_{j+2},$ and so on, until their budgets are
exhausted; the partial winner (bidder $j$) pays $v_{j+1}$ per unit up to a
budget of $b_{j+1},$ then pays $v_{j+2}$ per unit up to a budget of $b_{j+2},
$ and so on, until she spends $b_{j}-s.$\medskip 

Let us the denote the Sort-Cut revenue by $R^{S}\left( \mathbf{b},\mathbf{v}%
\right) $ (Note that $R^{S}\left( \mathbf{b},\mathbf{v}\right) =c^{\ast }$
where $X\left( c^{\ast },\left( \mathbf{b},\mathbf{v}\right) \right) =m$).
Next, we show that the revenue of Sort-Cut is nondecreasing in the budget
and value announcements of the bidders.

\begin{proposition}
\label{prop2}$R^{S}\left( \mathbf{b},\mathbf{v}\right) $ is nondecreasing in 
$\mathbf{b}$ and $\mathbf{v}$.
\end{proposition}

\begin{proof}
In the Appendix.
\end{proof}

\section{Truthfulness, Revenue, and Near Pareto Optimality}

\subsection{Truthfulness}

In this section, we show that Sort-Cut has good incentive properties. More
specifically, we show that no bidder benefits from understating her value or
budget. First we argue that three deviations that understate value or budget
or both are weakly dominated in ex-post equilibria. Then we consider two
other deviations that might potentially decrease revenue and argue that
either they are not reasonable or they result in higher revenue.

\begin{proposition}
\label{prop 3}For any bidder $i$ with types $\left( b_{i},v_{i}\right) ,$
bidding $\left( b_{i},v_{i}\right) $ weakly dominates bidding $\left(
b_{i},v_{i}^{-}\right) $ for $v_{i}^{-}<v_{i}.$
\end{proposition}

\begin{proof}
Consider any $\left( \mathbf{b}_{-i},\mathbf{v}_{-i}\right) .$ First of all,
if $i$ becomes a loser by bidding $\left( b_{i},v_{i}^{-}\right) ,$ her
utility cannot increase with this deviation. This is because losers'
utilities are zero, and by construction, a bidder with type $\left(
b_{i},v_{i}\right) $ achieves a nonnegative utility by bidding $\left(
b_{i},v_{i}\right) .$ We will look at the possible cases one by one.

\begin{itemize}
\item If $i$ is a loser by bidding $\left( b_{i},v_{i}\right) ,$ then she
will be a loser by bidding $\left( b_{i},v_{i}^{-}\right) $ (since the
pricing function gets better for the winners). Hence her utility cannot
increase by this deviation.

\item If $i$ is a partial winner by bidding $\left( b_{i},v_{i}\right) $ and
bidding $\left( b_{i},v_{i}^{-}\right) $ makes her a partial winner, then
she will have the same pricing function but she will be able to use less of
her budget (since the pricing function for full winners becomes better);
hence her utility cannot increase. Bidder $i$ cannot become a winner by
bidding $\left( b_{i},v_{i}^{-}\right) ,$ when she is a partial winner by
bidding $\left( b_{i},v_{i}\right) $.

\item If $i$ is a winner by bidding $\left( b_{i},v_{i}\right) $ and bidding 
$\left( b_{i},v_{i}^{-}\right) $ makes her a winner, her utility does not
change. This is because Sort-cut pricing ignores the value of winners in the
pricing calculation. If $i$ is a winner by bidding $\left(
b_{i},v_{i}\right) $ and bidding $\left( b_{i},v_{i}^{-}\right) $ makes her
a partial winner, then the original partial winner $j$ (with an unused
budget $s$) has to be a winner after the deviation. We argue that $i$'s
utility decreases. It is true that $i$ would get the items at a lower
per-unit price after the deviation, but at the same time she is using less
of her budget. The argument is that, by this deviation $i$ cannot get to
lower-priced items, and this follows from the fact that revenue of Sort-cut
cannot decrease after the deviation. More formally, let us denote the unused
budget of $i$ after the deviation by $s^{\prime }.$ We know that $s^{\prime
}\geq s$ (because revenue cannot increase). Bidder $i$'s utility difference
with the deviation can be shown to be nonpositive (where $\alpha $ and $c\ $%
are defined with respect to $\left( \mathbf{b},\mathbf{v}\right) $)%
\begin{eqnarray*}
&&\left( x\left( \alpha ^{c+s},b_{i}-s^{\prime }\right) v_{i}-\left(
b_{i}-s^{\prime }\right) \right) -\left( x\left( \alpha ^{c},b_{i}\right)
v_{i}-b_{i}\right) \\
&=&\left( x\left( \alpha ^{c+s},b_{i}-s^{\prime }\right) -x\left( \alpha
^{c},b_{i}\right) \right) v_{i}+s^{\prime } \\
&\leq &\left( x\left( \alpha ^{c+s^{\prime }},b_{i}-s^{\prime }\right)
-x\left( \alpha ^{c},b_{i}\right) \right) v_{i}+s^{\prime } \\
&=&\left( x\left( \alpha ^{c+s^{\prime }},b_{i}-s^{\prime }\right) -\left(
x\left( \alpha ^{c},s^{\prime }\right) +x\left( \alpha ^{c+s^{\prime
}},b_{i}-s^{\prime }\right) \right) \right) v_{i}+s^{\prime } \\
&=&s^{\prime }-x\left( \alpha ^{c},s^{\prime }\right) v_{i} \\
&\leq &s^{\prime }-\frac{s^{\prime }}{v_{i}}v_{i} \\
&=&0
\end{eqnarray*}%
where the first inequality follows from $s^{\prime }\geq s$ and the second
inequality follows from $\alpha ^{c}\left( y\right) \leq v_{i}.$
\end{itemize}
\end{proof}

\begin{proposition}
\label{prop 4}For any bidder $i$ with type $\left( b_{i},v_{i}\right) $,
bidding $\left( b_{i},v_{i}\right) $ weakly dominates bidding $\left(
b_{i}^{-},v_{i}\right) $ for $b_{i}^{-}<b_{i}.$
\end{proposition}

\begin{proof}
Consider any $\left( \mathbf{b}_{-i},\mathbf{v}_{-i}\right) $: First of all,
as in the previous proof, if $i$ becomes a loser by bidding $\left(
b_{i}^{-},v_{i}\right) ,$ her utility cannot increase with this deviation.
We look at the possible cases one by one.

\begin{itemize}
\item If $i$ is a loser by bidding $\left( b_{i},v_{i}\right) ,$ then she
will be a loser by bidding $\left( b_{i}^{-},v_{i}\right) $ (since the
pricing function gets better for the winners).

\item If $i$ is a partial winner by bidding $\left( b_{i},v_{i}\right) $ and
bidding $\left( b_{i}^{-},v_{i}\right) $ makes her a partial winner, then
she will have the same pricing function but will be able to use less of her
budget (since the pricing function for winners becomes better), hence her
utility cannot increase. Bidder $i$ cannot become a winner by bidding $%
\left( b_{i},v_{i}^{-}\right) ,$ when she is a partial winner by bidding $%
\left( b_{i},v_{i}\right) $.

\item If $i$ is a winner by bidding $\left( b_{i},v_{i}\right) $ and bidding 
$\left( b_{i}^{-},v_{i}\right) $ makes her a partial winner, then $i$ would
be worse off with this deviation. This is because (i) she is using less of
her budget, and (ii) her pricing got worse. If $i$ is a full winner by
bidding $\left( b_{i},v_{i}\right) $ and bidding $\left(
b_{i}^{-},v_{i}\right) $ leaves her a full winner, we can argue that her
utility decreases. It is true that $i$ may get the items at a lower per-unit
price after the deviation, but at the same time she is using less of her
budget. The argument is that by this deviation $i$ cannot get to
lower-priced items, which follows from the fact that the revenue of Sort-cut
cannot increase after the deviation. More formally, bidder $i$'s utility
difference with the deviation can be shown to be nonpositive as follows.
Here $\alpha $ and $c\ $are defined with respect to $\left( \mathbf{b},%
\mathbf{v}\right) $ and $c^{\prime }$ $(\leq c)$ is the Sort-cut revenue
after deviation.%
\begin{eqnarray*}
&&\left( x\left( \alpha ^{c^{\prime }+b_{i}-b_{i}^{-}},b_{i}^{-}\right)
v_{i}-b_{i}^{-}\right) -\left( x\left( \alpha ^{c},b_{i}\right)
v_{i}-b_{i}\right) \\
&=&\left( x\left( \alpha ^{c^{\prime }+b_{i}-b_{i}^{-}},b_{i}^{-}\right)
-x\left( \alpha ^{c},b_{i}\right) \right) v_{i}+b_{i}-b_{i}^{-} \\
&\leq &\left( x\left( \alpha ^{c+b_{i}-b_{i}^{-}},b_{i}^{-}\right) -x\left(
\alpha ^{c},b_{i}\right) \right) v_{i}+b_{i}-b_{i}^{-} \\
&=&\left( x\left( \alpha ^{c+b_{i}-b_{i}^{-}},b_{i}^{-}\right) -\left(
x\left( \alpha ^{c},b_{i}-b_{i}^{-}\right) +x\left( \alpha
^{c+b_{i}-b_{i}^{-}},b_{i}^{-}\right) \right) \right) v_{i}+b_{i}-b_{i}^{-}
\\
&=&b_{i}-b_{i}^{-}-x\left( \alpha ^{c},b_{i}-b_{i}^{-}\right) v_{i} \\
&\leq &b_{i}-b_{i}^{-}-\frac{b_{i}-b_{i}^{-}}{v_{i}}v_{i} \\
&=&0
\end{eqnarray*}%
where the first inequality follows from $c\geq c^{\prime }$ and the second
inequality follows from $\alpha ^{c}\left( y\right) \leq v_{i}$ for all $y$.%
\footnote{%
To see why bidder $i$'s pricing after the deviation is according to $\alpha
^{c^{\prime }+b_{i}-b_{i}^{-}},$ note that her pricing is according to $%
\alpha ^{c^{\prime }}$ according to types after the deviation, and this
translates to $\alpha ^{c^{\prime }+b_{i}-b_{i}^{-}}$ with the original
types.}
\end{itemize}
\end{proof}

Similarly, we can argue that bidding $\left( b_{i}^{-},v_{i}^{-}\right) $
for $b_{i}^{-}<b_{i}$ and $v_{i}^{\prime }<v_{i}^{-}$ is weakly dominated by
bidding $\left( b_{i},v_{i}\right) .$ This follows from the proofs above.
The two previous propositions imply that both $\left( b_{i}^{-},v_{i}\right) 
$ and $\left( b_{i},v_{i}^{-}\right) $ dominate $\left(
b_{i}^{-},v_{i}^{-}\right) $ when $b_{i}^{-}<b_{i}$ and $v_{i}^{-}<v_{i}.$
Applying either of them one more time, we have the following result.

\begin{proposition}
\label{prop 5}For any bidder $i$ with type $\left( b_{i},v_{i}\right) ,$
bidding $\left( b_{i},v_{i}\right) $ weakly dominates bidding $\left(
b_{i}^{-},v_{i}^{-}\right) $ for $b_{i}^{-}<b_{i}$ and $v_{i}^{-}<v_{i}.$
\end{proposition}

Propositions \ref{prop 3}, \ref{prop 4}, and \ref{prop 5} establish that
these-revenue decreasing deviations should not occur in equilibrium (they
are weakly dominated). There are two deviations, however, that may increase
or decrease the revenue. These deviations are \textquotedblleft understating
budget and overstating value\textquotedblright\ and \textquotedblleft
overstating budget and understating value.\textquotedblright\ We now show
that the former deviation is not reasonable in the sense that it could be a
best response only when the utility with that strategy is zero. Then we show
that the latter deviation could happen in equilibrium, yet whenever it is a
(strict) profitable deviation from truthful revelation, the revenue
increases with the deviation.

\begin{proposition}
\label{prop 6}For any bidder $i$ with type $\left( b_{i},v_{i}\right) ,$ for 
$b_{i}^{-}<b_{i}$ and $v_{i}^{+}>v_{i},$ bidding $\left(
b_{i}^{-},v_{i}^{+}\right) $ can never be in the set of best responses
unless bidder $i$'s utility in her best response is $0.$
\end{proposition}

\begin{proof}
Given $\left( \mathbf{b}_{-i},\mathbf{v}_{-i}\right) ,$ suppose that $\left(
b_{i}^{-},v_{i}^{+}\right) $ is a best response for $i$ where $%
b_{i}^{-}<b_{i}$ and $v_{i}^{+}>v_{i}.$ Since bidding $\left(
b_{i},v_{i}\right) $ would give nonnegative utility to bidder $i,$ the
utility by bidding $\left( b_{i}^{-},v_{i}^{+}\right) $ has to be
nonnegative. We claim that bidding $\left( b_{i},v_{i}^{+}\right) $ is a
better response than $\left( b_{i}^{-},v_{i}^{+}\right) ,$ and it is
strictly better when the utility by bidding $\left( b_{i},v_{i}^{+}\right) $
is strictly positive. This implies $\left( b_{i}^{-},v_{i}^{+}\right) $
could be a best response only when bidder $i$'s utility in her best response
is $0.$\medskip 

Suppose that utility by bidding $\left( b_{i}^{-},v_{i}^{+}\right) $ is
nonnegative, and consider the utility difference between bidding $\left(
b_{i}^{-},v_{i}^{+}\right) $ versus bidding $\left( b_{i},v_{i}^{+}\right) $%
. The utility difference is clearly zero if $i$ is a loser in both cases.
For all other cases, $i$ would be either a partial winner or a full winner
by bidding $\left( b_{i},v_{i}^{+}\right) .$ Then, we could see that bidding 
$\left( b_{i},v_{i}^{+}\right) $ gives a higher utility than bidding $\left(
b_{i}^{-},v_{i}^{+}\right) .$ The argument is the same as in the proof for
Proposition \ref{prop 4}: by bidding an extra budget of $b_{i}-b_{i}^{-}$
bidder $i$ can get extra items at a per-unit price lower than her value,
leading to a nonzero increase in her utility.
\end{proof}

In other words, we should not expect to see $\left(
b_{i}^{-},v_{i}^{+}\right) $ to be played, since it is worse than either $%
\left( b_{i},v_{i}\right) $ or $\left( b_{i},v_{i}^{+}\right) .$

\begin{proposition}
\label{prop 7}For any bidder $i$ with type $\left( b_{i},v_{i}\right) ,$ for 
$b_{i}^{+}>b_{i}$ and $v_{i}^{-}<v_{i},$ whenever bidding $\left(
b_{i}^{+},v_{i}^{-}\right) $ brings a higher utility to $i$ than bidding $%
\left( b_{i},v_{i}\right) ,$ the auctioneer's revenue with $\left(
b_{i}^{+},v_{i}^{-}\right) $ is not lower than the revenue with $\left(
b_{i},v_{i}\right) .$
\end{proposition}

\begin{proof}
Given $\left( \mathbf{b}_{-i},\mathbf{v}_{-i}\right) ,$ for some $%
b_{i}^{+}>b_{i}$ and $v_{i}^{-}<v_{i},$ suppose that $u_{i}\left( \left( 
\mathbf{b}_{-i},b_{i}^{+}\right) ,\left( \mathbf{v}_{-i},v_{i}^{-}\right)
\right) >u_{i}\left( \left( \mathbf{b}_{-i},b_{i}\right) ,\left( \mathbf{v}%
_{-i},v_{i}\right) \right) .$ Since bidder $i$ is budget constrained, she
will have to be a partial winner by bidding $\left(
b_{i}^{+},v_{i}^{-}\right) $ (if she is a full winner her utility would be $%
-C,$ and if she is a loser her utility would be $0$).

\begin{itemize}
\item If she is a loser by bidding $\left( b_{i},v_{i}\right) ,$ the
auctioneer's revenue clearly increases with $\left(
b_{i}^{+},v_{i}^{-}\right) .$ This is because $i$'s ranking with $v_{i}^{-}$
is not higher than with $v_{i},$ and so by deviating from $\left(
b_{i},v_{i}\right) $ to $\left( b_{i}^{+},v_{i}^{-}\right) ,$ all full
winners remain full winners and $i$ becomes a partial winner.

\item If she is a full winner by bidding $\left( b_{i},v_{i}\right) ,$ the
partial winner with $\left( b_{i},v_{i}\right) $ has to become a full winner
after $i$ deviates to $\left( b_{i}^{+},v_{i}^{-}\right) $. Otherwise, $i$
would be worse off by bidding $\left( b_{i}^{+},v_{i}^{-}\right) $ as she
will have a worse pricing function. In this case the revenue has to
increase. The argument is that, for this deviation to be beneficial, $i$ has
to get lower priced items after the deviation. For this to be the case, the
partial winner's unused budget before the deviation, plus $i$'s used budget
after the deviation, has to be greater than $i$'s budget $b_{i}.$ But in
this case, the revenue increases, since the new cut point is greater than
the old one.

\item If she is a partial winner by bidding $\left( b_{i},v_{i}\right) ,$ we
need to analyze two cases: (i) $i$'s ranking among the bidders is the same,
or (ii) $i$'s ranking is different. For (i), the pricing for $\left(
b_{i},v_{i}\right) $ and $\left( b_{i}^{+},v_{i}^{-}\right) $ are the same.
Since utility with $\left( b_{i}^{+},v_{i}^{-}\right) $ is more than utility
with $\left( b_{i},v_{i}\right) ,$ this means $i$ is using more of her
budget with $\left( b_{i}^{+},v_{i}^{-}\right) .$ Therefore the revenue
increases. For (ii), $i$'s ranking has to be worse with $\left(
b_{i}^{+},v_{i}^{-}\right) .$ Now, similar to the previous case, we argue
that total budget of \textquotedblleft new full winners\textquotedblright\
after the deviation plus the used budget of $i$ after deviation has to be
greater than $b_{i}.$ If that is not the case, $i$ cannot get to lower
prices.
\end{itemize}
\end{proof}

In the above propositions we argued that playing $\left(
b_{i}^{-},v_{i}\right) ,$ $\left( b_{i},v_{i}^{-}\right) $ and $\left(
b_{i}^{-},v_{i}^{-}\right) $ are not reasonable (they are dominated by $%
\left( b_{i},v_{i}\right) $); playing $\left( b_{i}^{-},v_{i}^{+}\right) $
is not reasonable in a weaker sense (it is dominated by a combination of $%
\left( b_{i},v_{i}\right) $ and $\left( b_{i},v_{i}^{+}\right) $); also,
playing $\left( b_{i}^{+},v_{i}^{-}\right) $ is reasonable only when it is
done by a winner, who becomes a partial winner after deviation and increases
the overall revenue. We call the equilibria in which the strategies satisfy
these conditions a \emph{refined equilibrium. }

\begin{definition}
A \emph{refined equilibrium} is an equilibrium of Sort-Cut where for all
bidders $i$, bidder $i$ does not play $\left( b_{i}^{-},v_{i}\right) $, $%
\left( b_{i},v_{i}^{-}\right) ,$ $\left( b_{i}^{-},v_{i}^{-}\right) $, or $%
\left( b_{i}^{-},v_{i}^{+}\right) .$ Moreover, a bidder $i$ plays $\left(
b_{i}^{+},v_{i}^{-}\right) $ only when $u_{i}\left( \left( \mathbf{b}%
_{-i},b_{i}^{+}\right) ,\left( \mathbf{v}_{-i},v_{i}^{-}\right) \right)
>u_{i}\left( \left( \mathbf{b}_{-i},b_{i}\right) ,\left( \mathbf{v}%
_{-i},v_{i}\right) \right) .$
\end{definition}

The refined equilibrium refinement of Sort-Cut's ex-post Nash equilibria is
in the spirit of refinement of considering weakly undominated strategies: $%
\left( b^{-},v\right) $, $\left( b,v^{-}\right) ,$ and $\left(
b^{-},v^{-}\right) $ are weakly dominated; $\left( b^{-},v^{+}\right) $ is
dominated by a combination of two strategies; and our refinement requires $%
\left( b^{+},v^{-}\right) $ to be played when it is better than $\left(
b,v\right) .$ In a refined equilibrium, bidders never understate their
budgets, and they understate their values only when they also simultaneously
overstate their budgets, making them better off than their truthful
announcements. Recall that when $u_{i}\left( \left( \mathbf{b}%
_{-i},b_{i}^{+}\right) ,\left( \mathbf{v}_{-i},v_{i}^{-}\right) \right)
>u_{i}\left( \left( \mathbf{b}_{-i},b_{i}\right) ,\left( \mathbf{v}%
_{-i},v_{i}\right) \right) ,$ $\left( b_{i}^{+},v_{i}^{-}\right) $ makes $i$
a partial winner after the deviation and revenue is higher with $\left(
b_{i}^{+},v_{i}^{-}\right) $ than with $\left( b_{i},v_{i}\right) $.

\subsection{Revenue}

There are eight possible kinds of deviations from truthful revelation, $%
\left( b_{i},v_{i}\right) .$ Five of them are discussed in the definition of
refined equilibrium, and the remaining three of them, namely $\left(
b_{i},v_{i}^{+}\right) ,$ $\left( b_{i}^{+},v_{i}\right) ,$ and $\left(
b_{i}^{+},v_{i}^{+}\right) ,$ can only increase the revenue by Proposition~%
\ref{prop2}. Hence we have the following result.

\begin{theorem}
\label{prop 8}In a refined equilibrium of Sort-Cut, revenue is bounded below
by the revenue of Sort-Cut with truthful revelations.
\end{theorem}

\begin{proof}
Consider any refined equilibrium of Sort-Cut. Let $b_{i}^{-}$ and $v_{i}^{-}$
denote understating the types, and $b_{i}^{+}$ and $v_{i}^{+}$ denote
overstating the types (with respect to true types). We know that $\left(
b_{i}^{-},v_{i}\right) $, $\left( b_{i},v_{i}^{-}\right) ,$ $\left(
b_{i}^{-},v_{i}^{-}\right) ,$ or $\left( b_{i}^{-},v_{i}^{+}\right) $ do not
occur. Additionally, $\left( b_{i}^{+},v_{i}^{-}\right) $ could only occur
for the current cut-point bidder, and by Proposition \ref{prop 7}, if we
change it back to $\left( b_{i},v_{i}\right) ,$ revenue cannot increase.
Finally, the rest of the bidders are either bidding truthfully or using $%
\left( b_{i},v_{i}^{+}\right) ,$ $\left( b_{i}^{+},v_{i}\right) $, or $%
\left( b_{i}^{+},v_{i}^{+}\right) .$ In any case, changing their bid to
their truthful values cannot increase the revenue. Therefore revenue in a
refined equilibrium of Sort-Cut is not smaller than revenue of Sort-Cut with
truthful revelations.
\end{proof}

\subsection{Near Pareto Optimality}

Among different efficiency concepts that could be considered, we consider
that of Pareto optimality: we say that an allocation is Pareto optimal if
there is no other allocation in which all players (including the auctioneer)
are better off and at least one player is strictly better off.\footnote{%
Maximizing social welfare dictates all items to be allocated to the bidder
with the highest value, even if this bidder has a very small budget. We
follow Dobzinski et al. (2008) and consider Pareto optimality as the
appropriate efficiency concept.} In this setup, Dobzinski et al. (2008) has
shown that Pareto optimality is equivalent to a \textquotedblleft no
trade\textquotedblright\ condition: an allocation is Pareto efficient if (a)
all units are sold and (b) a player get a non-zero allocation only if all
higher-value players exhaust their budgets. In other words, an allocation is
Pareto optimal when, given the \emph{true value} of the partial winner,
winners and losers are ordered in the right way: all winners have higher
values and all losers have lower values.\medskip 

The previous subsection demonstrated that Sort-Cut has good revenue
properties. The following result pertains to the efficiency (near Pareto
optimality) of the equilibria of Sort-Cut. It shows that in any ex-post Nash
equilibrium of Sort-Cut, the full winners and losers are ordered in the
right way given the \emph{announced value} of the partial winner.

\begin{theorem}
\label{prop 9} Consider any ex-post Nash equilibrium of Sort-Cut where $%
v_{j} $ is the announced value of the partial winner $j.$ Every bidder $%
i\neq j$ who has a true value $v_{i}^{T}>v_{j}$ is a full winner, and every
bidder $i\neq j$ who has a true value $v_{i}^{T}<v_{j}$ is a loser in this
equilibrium of Sort-Cut.
\end{theorem}

\begin{proof}
First, consider a bidder $i$ whose value is $v_{i}^{T}>v_{j}.$ We prove that
she must be a full winner in equilibrium. Assume for the sake of
contradiction that bidder $i$ is a loser, so her utility is zero. If she
deviates and bids $v_{j}+\varepsilon $ (for $0<\varepsilon <v_{i}^{T}-v_{j}$%
) and her true budget, she will become either a full winner or the cut-point
bidder (otherwise revenue of Sort-cut will decrease with this deviation,
which is not possible because of Proposition \ref{prop2}). Obviously her
utility becomes strictly positive with this deviation (her price per unit is
at most $v_{j}$). We thus reach the necessary contradiction to her
individual rationality.\medskip 

Now consider a bidder $i$ whose value is $v_{i}^{T}<v_{j}.$ Assume for the
sake of contradiction that bidder $i$ is a full winner. If $b_{i}$ is
smaller than the unused budget of the cut-point bidder ($s$), then she gets
all items at a per-unit price $v_{j},$ and hence she obtains a negative
utility. If this is the case, she would be better off announcing her true
valuations to guarantee a nonnegative payoff. If $b_{i}>s,$ then we argue
that $i$ would be better of by deviating to $(v_{j}-\varepsilon ,b_{i})$ for
small enough $\varepsilon .$ Let us first look at the limiting case in which 
$i$ deviates to $(v_{j},b_{i})$ and becomes the cut-point bidder. After this
deviation, the unused budget of $i$ would be exactly $s.$ The allocation of
original full winners will not change; bidder $j$ will be getting $\frac{s}{%
v_{j}}$ more items by paying $s$ more and bidder $i$ will be getting $\frac{s%
}{v_{j}}$ less items by paying $s$ less. Therefore, bidder $i$'s utility
increases by $\frac{s}{v_{j}}\left( v_{j}-v_{i}^{T}\right) >0$ (in a sense
by this deviation, bidder $i$ is selling $\frac{s}{v_{j}}$ units of the
items to bidder $j$ at the per-unit price of $v_{j}$). By deviating to $%
(v_{j}-\varepsilon ,b_{i}),$ the original full winners' allocations would
slightly increase; therefore bidder $i$'s utility increase will be slightly
smaller than $\frac{s}{v_{j}}\left( v_{j}-v_{i}^{T}\right) .$\footnote{%
There is an implicit continuity assumption here. However, it is not
difficult to show that utilities of the bidders are continuous in type
announcements.} But for small enough $\varepsilon ,$ it will always be
positive, leading again to a contradiction.\medskip 
\end{proof}

This theorem establishes that given equilibrium cut-point value, all winners
and losers will be rightly placed. But since the cut-point bidder may be
misplaced, this does not imply full Pareto optimality. Consider the
following example.

\begin{example}
There are 2 units of the item to be sold, and there are four bidders with
budget-value pairs $\left( 18,19\right) ,\left( 1,9\right) ,\left( \frac{17}{%
9},8\right) ,$ and $\left( 10,1\right) .$ For this setup, it can be
confirmed that bidders announcing their types (budget,value) as $\left(
18,19\right) ,\left( 1,9\right) ,\left( 36,18\right) ,$ and $\left(
10,1\right) $ constitute an ex-post equilibrium of Sort-Cut. In this
equilibrium, bidder 3 overstates her value and budget and becomes the
partial winner. Although the full winners and the losers are rightly ranked
according to the announced value of the partial winner, the allocation is
not Pareto optimal. Bidder 3 gets a positive allocation even though bidder 2
has a higher value and zero allocation.\medskip 

As an aside, note that the revenue of Sort-Cut in this ex-post equilibrium
is $18$ while the revenue with truthful types is $18+\frac{17}{9}$.
\end{example}

\section{Market Clearing Price Mechanism and Sort-Cut}

In this section we compare Sort-Cut with the well-known \emph{Market
Clearing Price Mechanism (MCPM). }MCPM is a mechanism that sells $m$ items
to all interested bidders at a fixed price. That is, in MCPM all items are
sold $p$ dollars per unit and all bidders whose values are strictly greater
than $p$ spend all their budgets to buy these items (the bidders with values
equal to $p$ could be partially spending their budgets to clear the supply).

\begin{definition}
The Market Clearing Price Mechanism is an $m$-Procedure cut mechanism with
fixed pricing rule, $\alpha \left( y\right) =p^{\ast }$ for all $y\geq 0$
where $p^{\ast }$ satisfies $v_{j}\geq p^{\ast }>v_{j-1}.$
\end{definition}

One can easily argue that there will be a unique $p^{\ast }.$ Consider a
fixed pricing rule $\alpha \left( y\right) =p^{\ast }$ that satisfies the
above definition. Then for any fixed pricing rule $\alpha \left( y\right) =p$
with $p>p^{\ast },$ we have $p>v_{j}$ (for $v_{j}$ defined by $\alpha \left(
y\right) =p$); and for any $p<p^{\ast },$ we have $p\leq v_{j-1}$ (for $%
v_{j-1}$ defined by $\alpha \left( y\right) =p$).\medskip 

Although it is a natural mechanism, as we demonstrate below, MCPM lacks good
incentive properties.

\begin{proposition}
\label{prop 10}Under the MCPM, overstating budget or value is weakly
dominated by bidding true types, i.e, for bidder $i$ with type $\left(
b_{i},v_{i}\right) $, announcing $\left( b_{i}^{+},v_{i}\right) ,\left(
b_{i},v_{i}^{+}\right) ,$ and $\left( b_{i}^{+},v_{i}^{+}\right) $ are all
weakly dominated by $\left( b_{i},v_{i}\right) $.
\end{proposition}

\begin{proof}
Consider bidder $i$ with type $\left( b_{i},v_{i}\right) $ who announces her
type truthfully.

\begin{itemize}
\item If she is a full winner, she is indifferent to announcing $v_{i}^{+}$
and would be strictly worse off by announcing $b_{i}^{+}$ (she would either
get negative payoff by staying a full winner or will get zero utility by
becoming a partial winner or a loser).

\item If she is a partial winner, since overstating her value or budget can
only increase the market clearing price $p^{\ast },$ she never can obtain a
strictly positive payoff by deviating to $v_{i}^{+}$ or $b_{i}^{+}.$

\item If she is a loser, by overstating her value or budget, she may become
a winner, but the market clearing price after deviation is going to be
greater than the previous market clearing price and hence greater than her
value.
\end{itemize}
\end{proof}

However, understating her value or budget in general can be beneficial.
Consider the following example.

\begin{example}
Consider two bidders with (budget,value) pairs $\left( 16,10\right) $ and $%
\left( 8,9\right) $ where the supply is $m=3.$ Under truthful report of
types, the market clearing price is $p^{\ast }=8.$ However, if bidder 1
understates her value as $7,$ the new market clearing price will be $7$ with
the first bidder spending $13$ of her budget for an allocation of $\frac{13}{%
7}$ units. Her new payoff is $\left( 10-7\right) \frac{13}{7}\cong 5.57$
versus $\left( 10-8\right) 2=4,$ which shows that her understatement of
value is a profitable deviation. Similarly, if bidder $1$ understates her
budget as $10,$ the new market clearing price will be $6$ with the first
bidder spending $10$ of her budget for an allocation of $\frac{10}{6}$
units. Her new payoff is $\left( 10-6\right) \frac{10}{6}\cong 6.67,$ which
shows that her understatement of budget is a profitable deviation. In fact,
for this example an ex-post equilibrium is when bidders announce their types
as $\left( \frac{2700}{361},10\right) ,$ $\left( \frac{2430}{361},9\right) ,$
which brings only a revenue of $14.21$ (in comparison with truthful revenue
of $24.$)
\end{example}

The above discussion illustrates that the revenue from an (undominated)
ex-post equilibrium of MCPM is bounded above by the revenue of MCPM with
truthful revelations. Next, we obtain a lower bound for the revenue of
Sort-Cut. For any announcements $\left( \mathbf{b},\mathbf{v}\right) ,$ we
show that the revenue difference between MCPM and Sort-Cut is at most equal
to the maximum budget of the players. For the same announcement of the
types, since Sort-Cut's pricing function for the winners decreases in
winners' budgets, whereas MCPM's pricing is constant, MCPM's revenue would
be higher than the revenue of Sort-Cut. However, the following proposition
shows that the difference in revenues is bounded above by the maximum of the
winners' budgets. Let $R^{M}\left( \mathbf{b},\mathbf{v}\right) $ denote
MCPM's revenue and $b_{\max }$ denote the maximum budget of the bidders.%
\footnote{$b_{\max }$ can also be taken as the maximum budget of the full
winners.}

\begin{proposition}
\label{prop 11}For any announcements $\left( \mathbf{b},\mathbf{v}\right) ,$ 
$R^{M}\left( \mathbf{b},\mathbf{v}\right) -R^{S}\left( \mathbf{b},\mathbf{v}%
\right) \leq b_{\max }$.
\end{proposition}

\begin{proof}
Given $\left( \mathbf{b},\mathbf{v}\right) ,$ let Sort-Cut's cut point be
denoted by $c^{\ast },$ and let MCPM's cut point (fixed price) be denoted by 
$c^{\ast \ast }$. We argue that $c^{\ast \ast }-c^{\ast }\leq b_{\max }.$ By
the definition of MCPM, $c^{\ast \ast }=m\times p^{\ast }$ where $p^{\ast }$
satisfies $v_{j}\geq p^{\ast }>v_{j-1}$ and $v_{j}$ is the partial winner in
MCPM. Since $c^{\ast }\leq c^{\ast \ast },$ $j$ cannot be a full winner in
Sort-Cut. If she is a partial winner, then $c^{\ast \ast }-c^{\ast }\leq
b_{j}\leq b_{\max }$ holds since the difference between $c^{\ast \ast }$ and 
$c^{\ast }$ is smaller than $b_{j}.$ If $j$ is a loser in Sort-Cut, then we
argue as follows. At least one of the winners of Sort-cut has to pay at most 
$p^{\ast }$ per unit (otherwise the revenue of Sort-Cut has to be greater
than $c^{\ast \ast }$). Now, this bidder's budget has to be greater than $%
c^{\ast \ast }-c^{\ast },$ because otherwise her price per unit cannot be
smaller than $p^{\ast }.$ Hence, $c^{\ast \ast }-c^{\ast }\leq b_{\max
}.\medskip $
\end{proof}

Proposition \ref{prop 11}, together with proposition \ref{prop 8},
establishes the following result.

\begin{theorem}
Let us denote the revenue of MCPM with the truthful revelation of types by $%
R^{\ast }$. Then the revenue of any refined equilibrium of Sort-Cut is not
lower than $R^{\ast }-b_{\max }.$
\end{theorem}

Unlike Sort-Cut, we next show an example where MCPM obtains a revenue that
is an order of magnitude (as the number of bidders) lower than $R^{\ast }$.

\begin{example}
Consider two types of bidders with budget, value pairs $%
(b_{0},v_{0})=(16,18) $ and $(b_{1},v_{1})=(8,9)$; our basic example has one
bidder of each type with a supply of $m=3$ units. Under truthful reports of
budgets and values, the market-clearing price is $p=8$. Let us look for an
ex-post equilibrium, in which the announcements are $\left( a_{0},18\right) $
and $\left( a_{1},9\right) .$ The pair of values $a_{0}$ and $a_{1}$ solve
the optimization problems of $\max (v_{i}-p)\frac{a_{i}}{p}$ for $i=0,1$
where $p $ is the market clearing price for the given announcements and
supply. In our case $p=\frac{a_{0}+a_{1}}{3}$.$\medskip $

Thus the optimization problem becomes $\max f(a_{i})=\frac{3v_{i}a_{i}}{%
a_{i}+a_{1-i}}-a_{i}$. Taking derivatives, we get $f^{\prime }(a_{i})=\frac{%
3v_{i}a_{1-i}}{(a_{1}+a_{1-i})^{2}}-1$, with $f"(a_{i})<0$. Solving the pair
of first-order equations by setting $f^{\prime }(a_{0})=f^{\prime }(a_{1})=0$%
, we get $a_{1}=6$ and $a_{2}=12$ for a market clearing price of 6. The
total revenue of this equilibrium is therefore $18$ compared to $R^{\ast
}=24 $.$\medskip $

If we now scale the example to have N bidders of each type and a supply of $%
3N$, we may assume that all the optimal budget announcements of each type of
bidder are the same by symmetry. The clearing price stays unchanged at $p=%
\frac{N(a_{0}+a_{1})}{3N}=\frac{a_{0}+a_{1}}{3}$. The optimization problem
for determining each $a_{i}$ remains identical, giving the same solutions as
before.\medskip 

However, the revenue now is $16N$ compared to $R^{\ast }=24N$ and is thus a
whole third less than $R^{\ast }$, while by Theorem 3, Sort-Cut's revenue in
a refined equilibrium is not smaller than $24N-16.$
\end{example}

\section{Conclusion and Discussion}

In this paper, we have introduced a mechanism to sell $m$ divisible units to
a set of bidders with budget constraints. In this practically important
setting, where a mechanism that is simultaneously truthful and Pareto
optimal is precluded, our mechanism, Sort-Cut, achieves good incentive,
revenue, and efficiency properties. Specifically, in Sort-Cut, (i) there are
profitable deviations from truthful revelations of types, but these can only
happen in a revenue-increasing way; (ii) in a refined ex-post equilibrium,
the revenue of Sort-Cut is bounded below by $R^{\ast }-b_{\max },$ and (iii)
the equilibrium allocation is \emph{nearly Pareto efficient }in the sense
that full winners and losers are ordered in the right way given the
announced value of the partial winner. We then compare Sort-Cut to a
well-known mechanism, Market Clearing\ Price Mechanism (MCPM). We show that
in MCPM, (i) revenue increasing deviations are dominated, and (ii) the
revenue can be much smaller than $R^{\ast }-b_{\max }.$\medskip 

There are many ways our work can be generalized. In the context of online
advertisement auctions, our model can be interpreted as \textquotedblleft
there is a \emph{single} sponsored link that gets $m$ clicks a day and there
are $n$ advertisers.\textquotedblright\ However, in reality, there are many
sponsored links. In generalized second-price auctions studied by Edelman et
al. (2007), the winner of the best item (first sponsored link) is charged
the bid of the second-best item, the winner of the second-best item is
charged the bid of the third-best item, and so on. In this environment there
are no budget constraints and the second-highest bid is always the
competitor of the highest value. The idea of Sort-Cut can be applied in this
setup with budget constraints. More specifically, it would be interesting to
consider a model in which there are budget-constrained bidders and multiple
slots available for a query (in which an advertiser cannot appear in more
than one slot per query).\medskip 

In our model, we consider a setting of hard budget constraints in which the
bidders cannot spend more than their budgets. Extending our results to a
soft-budget problem in which bidders are able to finance further budgets at
some cost is a promising direction. One can model this kind of soft-budget
constraint as specifying value per-click up to some budget, then specifying
a smaller value per-click up to some other extra budget, and so on. By
replicating a bidder into as many copies as the number of pieces in her
value/budget function, and allowing them all to participate in our
mechanism, it seems reasonable that we may preserve some of the desirable
properties of Sort-Cut.\medskip 

One very important extension is to consider the environment of multi-item
auctions with budget constraints. Again consider advertisement department of
a computer manufacturer, who this time wants to appear in a search engine's
queries for \textquotedblleft laptops\textquotedblright\ and
\textquotedblleft desktops.\textquotedblright\ This advertisement department
might have a total budget to allocate between all online ads and their
per-click values for different items might be different. For instance one
firm might have higher per-click values for desktops, but lower per-click
values of laptops, as compared to a second firm. Designing an allocation and
pricing rule that would have good efficiency and revenue properties for this
setup is very challenging. Devanur et al. (2002) provided an algorithm for
finding \textquotedblleft market clearing prices.\textquotedblright\ This
mechanism, however, lacks good incentive properties. Bidders would have an
incentive to understate their budgets, thereby decreasing the prices. The
extension of Sort-Cut to this setting is not straightforward, since how
bidders will want to split the budgets between different items would depend
on the pricing rule of each of these items, which in turn depends on the
budget splits. Bidders' effective valuations for different goods are given
by the ratios of \textquotedblleft per-click values and the average
prices\textquotedblright\ of different items. This multi-item extension
seems to be the most important yet also the most challenging extension of
our model.

\bigskip

\section{Appendix}

\subsection{Proof of Proposition \protect\ref{prop1}}

First, note that $x\left( \alpha ^{c},b\right) $ is weakly increasing in $c:$
since $\alpha $ is nonincreasing, for $c^{\prime }\geq c\geq 0$, we have $%
\alpha ^{c^{\prime }}\left( y\right) =\alpha \left( y+c^{\prime }\right)
\leq \alpha \left( y+c\right) =\alpha ^{c}\left( y\right) ,$ and hence%
\begin{equation*}
x\left( \alpha ^{c^{\prime }},b\right) =\int_{0}^{b}\frac{1}{\alpha
^{c^{\prime }}\left( y\right) }dy\geq \int_{0}^{b}\frac{1}{\alpha ^{c}\left(
y\right) }dy=x\left( \alpha ^{c},b\right) .
\end{equation*}

Also, obviously $x\left( \alpha ^{c},b\right) $ is strictly increasing in $b.
$\medskip 

Now, we can show that $X\left( c,\left( \mathbf{b},\mathbf{v}\right) \right) 
$ is strictly increasing in $c.$ Consider $c^{\prime }>c\geq 0;$ we have 
\begin{equation*}
X\left( c,\left( \mathbf{b},\mathbf{v}\right) \right) =\left(
\sum\nolimits_{i=1}^{j-1}x\left( \alpha ^{c},b_{i}\right) \right) +x\left(
\alpha ^{c+s},b_{j}-s\right) 
\end{equation*}%
where $j$ satisfies $c\leq \sum\nolimits_{i=1}^{j}b_{i}$ and $%
c>\sum\nolimits_{i=1}^{j-1}b_{i}$ (and $s=\sum\nolimits_{i=1}^{j}b_{i}-c$).
For $c^{\prime }>c,$ we can have one of the two cases: either $j$ is the
same or $j$ is bigger.\medskip 

If $j$ is bigger, then we have 
\begin{eqnarray*}
X\left( c^{\prime },\left( \mathbf{b},\mathbf{v}\right) \right)
&>&\sum\nolimits_{i=1}^{j}x\left( \alpha ^{c^{\prime }},b_{i}\right) \\
&\geq &\left( \sum\nolimits_{i=1}^{j-1}x\left( \alpha ^{c},b_{i}\right)
\right) +x\left( \alpha ^{c^{\prime }},b_{j}\right) >X\left( c,\left( 
\mathbf{b},\mathbf{v}\right) \right)
\end{eqnarray*}

This is because $x\left( \alpha ^{c^{\prime }},b_{i}\right) \geq x\left(
\alpha ^{c},b_{i}\right) $ for all $i=1,..,j-1$ and $x\left( \alpha
^{c^{\prime }},b_{j}\right) >x\left( \alpha ^{c+s},b_{j}-s\right) $ since $%
c^{\prime }>c+s.$\medskip 

If $j$ is the same (if $c^{\prime }<c+s$), then we have%
\begin{eqnarray*}
X\left( c^{\prime },\left( \mathbf{b},\mathbf{v}\right) \right)  &=&\left(
\sum\nolimits_{i=1}^{j-1}x\left( \alpha ^{c^{\prime }},b_{i}\right) \right)
+x\left( \alpha ^{c^{\prime }+s^{\prime }},b_{j}-s^{\prime }\right)  \\
&>&\left( \sum\nolimits_{i=1}^{j-1}x\left( \alpha ^{c},b_{i}\right) \right)
+x\left( \alpha ^{c+s},b_{j}-s\right) =X\left( c,\left( \mathbf{b},\mathbf{v}%
\right) \right) 
\end{eqnarray*}%
where $s^{\prime }=\sum\nolimits_{i=1}^{j}b_{i}-c^{\prime }<s.$ This is
because $x\left( \alpha ^{c^{\prime }},b_{i}\right) \geq x\left( \alpha
^{c},b_{i}\right) $ for all $i=1,..,j-1$ and $x\left( \alpha ^{c^{\prime
}+s^{\prime }},b_{j}-s^{\prime }\right) >x\left( \alpha
^{c+s},b_{j}-s\right) $ since $c^{\prime }+s^{\prime }=c+s$ and $%
b_{j}-s^{\prime }>b_{j}-s.$\medskip 

Next, we show that $X\left( c,\left( \mathbf{b},\mathbf{v}\right) \right) $
is continuous in $c.$ By definition, $x\left( \alpha ^{c},b\right) $ is
continuous in $c$ and $b$ (this is because $x\left( \alpha ^{c},b\right)
=\int_{0}^{b}\frac{1}{\alpha \left( y+c\right) }dy$ and is continuous in $c$
and $b$ even when $\alpha $ is not a continuous function). Moreover,%
\begin{equation*}
X\left( c,\left( \mathbf{b},\mathbf{v}\right) \right) =\left(
\sum\nolimits_{i=1}^{j-1}x\left( \alpha ^{c},b_{i}\right) \right) +x\left(
\alpha ^{c+s},b_{j}-s\right) .
\end{equation*}

If $c$ increases from $c$ to $c+\varepsilon ,$ $j$ changes only when $s=0.$
If $s\neq 0,$ then $X\left( c,\left( \mathbf{b},\mathbf{v}\right) \right) $
is obviously continuous in $c$ as all of the terms in the summation are
continuous in $c.$ If $s=0,$ then 
\begin{equation*}
X\left( c+\varepsilon ,\left( \mathbf{b},\mathbf{v}\right) \right) =\left(
\sum\nolimits_{i=1}^{j}x\left( \alpha ^{c+\varepsilon },b_{i}\right) \right)
+x\left( \alpha ^{c+\varepsilon +s^{\prime }},b_{j+1}-s^{\prime }\right)
\end{equation*}%
and this goes to $X\left( c,\left( \mathbf{b},\mathbf{v}\right) \right) $ as 
$\varepsilon $ goes to zero. This is because $\sum\nolimits_{i=1}^{j}x\left(
\alpha ^{c+\varepsilon },b_{i}\right) \rightarrow
\sum\nolimits_{i=1}^{j}x\left( \alpha ^{c},b_{i}\right) =X\left( c,\left( 
\mathbf{b},\mathbf{v}\right) \right) $ and $x\left( \alpha ^{c+\varepsilon
+s^{\prime }},b_{j+1}-s^{\prime }\right) \rightarrow 0$ since $s^{\prime
}\rightarrow b_{j+1}.$

\subsection{Proof of Proposition \protect\ref{prop2}}

Consider bidder $i$ with announced type $\left( b_{i},v_{i}\right) .$

\begin{itemize}
\item First, we show that revenue is nondecreasing in budgets. Consider
bidder $i$ who decreases her budget to $b_{i}^{-}<b_{i}.$ We show that
revenue cannot increase with this deviation.

\begin{itemize}
\item If bidder $i$ was originally a loser by announcing $\left(
b_{i},v_{i}\right) ,$ then she cannot become a winner or partial winner by
deviating to $b_{i}^{-}<b_{i}.$ This is because by this deviation, the
pricing function for everybody becomes better and winners pay less per unit.
Therefore the revenue cannot increase.\medskip 

\item Next, consider bidder $i$ who is a partial winner by bidding $b_{i}.$
If bidder $i$ deviates to $b_{i}^{-}$ and becomes a loser, then the revenue
has to decrease since the set of losers becomes larger with this deviation.
If she deviates to $b_{i}^{-}$ and remains a partial winner, since all
winners' pricing gets better, the revenue has to decrease. If she deviates
to $b_{i}^{-},$ she cannot become a full winner. If this were the case, the
pricing function for every (full or partial) winner gets better, and then
the total number of units allocated will be greater than $m.$\medskip 

\item Lastly, consider bidder $i$ who is originally a winner by announcing $%
\left( b_{i},v_{i}\right) $. If she deviates to $b_{i}^{-}$ and if she
becomes a loser or a partial winner after the deviation, then the revenue
clearly decreases. This is because the set of full winners before the
deviation is a strict superset of the set of full winners after the
deviation. Now consider the case where bidder $i$ deviates to $b_{i}^{-}$
and remains a winner. Let us denote $b_{i}-b_{i}^{-}$ by $\Delta .$ Suppose
that the initial cut point is $c$ and the new cut point after the deviation
is $c^{\prime }.$ Let $\alpha $ be the $n$-piece step function defined by $%
\left( \mathbf{b},\mathbf{v}\right) .$ Note that the initial revenue is $c$
and the new revenue is $c^{\prime }.$ We will show that $c\geq c^{\prime }.$%
\medskip 

Since $i$ has understated her budget, there will be a shortage of demand and
the pricing of all original winners will be better. Therefore, with this
deviation, all original winners except $i$ will be allocated (weakly) more
units of the object. Assume for contradiction that $c^{\prime }>c.$ This
means that there will be new winners who use an extra budget strictly
greater than $\Delta ,$ say $\Delta ^{\prime }.$ We now argue that the extra
units allocated to these new winners have to be greater than the number of
units $i$ is giving up with the deviation. Extra units allocated to new
winners are priced at the values starting from the new cut point $c+\Delta
^{\prime }$ (according to $\left( \mathbf{b},\mathbf{v}\right) $) and the
total budget used is $\Delta ^{\prime }$. The number of units $i$ is giving
up are priced at the values in the range of $c$ to $c+\Delta <c+\Delta
^{\prime }$ and the total budget used is $\Delta .$ Since extra units are
given with higher budget ($\Delta ^{\prime }>\Delta $) and lesser prices ($%
c+\Delta <c+\Delta ^{\prime }$) than the units given up, we conclude that
with the assumption $c^{\prime }>c,$ the total number of units allocated has
to be strictly greater than $m,$ which is a contradiction.\medskip 

We can present this argument more formally. Consider the case in which $%
\Delta $ is small enough so that the original partial winner $j$ remains a
partial winner. All full winners $k\neq i$ with $k<j$ will be allocated more
items since $j$ will be using more of her budget after the deviation. Let us
consider the difference between the total amounts allocated to bidders $i$
and $j$ before and after the deviation. Bidder $i$'s allocation is decreased
by%
\begin{equation*}
A\equiv x\left( \alpha ^{c},b\right) -x\left( \alpha ^{c+\Delta ^{\prime
}},b-\Delta \right) 
\end{equation*}%
since%
\begin{equation*}
x\left( \alpha ^{c+\Delta ^{\prime }},b-\Delta \right) >x\left( \alpha
^{c+\Delta ^{\prime }},b-\Delta ^{\prime }\right) .
\end{equation*}%
We have%
\begin{eqnarray*}
A &<&x\left( \alpha ^{c},b\right) -x\left( \alpha ^{c+\Delta ^{\prime
}},b-\Delta ^{\prime }\right)  \\
&=&x\left( \alpha ^{c},\Delta ^{\prime }\right) .
\end{eqnarray*}%
On the other hand, bidder $j$'s allocation is increased by%
\begin{eqnarray*}
B &\equiv &x\left( \alpha ^{c+s},b_{j}-s+\Delta ^{\prime }\right) -x\left(
\alpha ^{c+s},b_{j}-s\right)  \\
&=&x\left( \alpha ^{c+b_{j}},\Delta ^{\prime }\right) 
\end{eqnarray*}%
since%
\begin{equation*}
x\left( \alpha ^{c+b_{j}},\Delta ^{\prime }\right) \geq x\left( \alpha
^{c},\Delta ^{\prime }\right) 
\end{equation*}%
we conclude $B>A.$ The argument for the case when the deviation results in a
change of the partial winner is very similar but not illuminating. Thus, the
total number of units allocated has to increase after the deviation.
\end{itemize}

\item Now, we show that revenue is increasing in values. Consider bidder $i$
who increases her value to $v_{i}^{+}>v_{i}.$ We show that revenue cannot
decrease with this deviation.

\begin{itemize}
\item First, if bidder $i$ is a winner by bidding $\left( b_{i},v_{i}\right) 
$ and she deviates to $v_{i}^{+}>v_{i},$ then she remains a winner after the
deviation, and the revenue does not change. This is because Sort-Cut's
allocation and pricing rule is invariant to full winners' values (so long as
they remain full winners).\medskip 

\item Second, consider a bidder $i$ who is a loser by bidding $\left(
b_{i},v_{i}\right) $ and deviates to $v_{i}^{+}>v_{i}.$ If she remains a
loser after the deviation, since the pricing function for winners gets
worse, the revenue has to increase. Let us now consider the deviation which
makes $i$ a partial winner. If the partial winner becomes a full winner
after the deviation ($v_{i}^{+}<v_{j}$ where $j$ is the original partial
winner), the revenue obviously increases with the deviation, since the
cut-point has increased.\medskip 

Let us consider the case in which $v_{i}^{+}>v_{j}$; $i$ becomes a partial
winner and $j$ becomes a loser after the deviation. Assume for contradiction
that the revenue decreases with the deviation. If this is the case, it can
be seen that the pricing function for all winners becomes worse after the
deviation (total unspent budget of price setters with $v_{k}\geq v_{i}$
becomes greater and some of the values increase). Hence all full winners
will be allocated less units of items after the deviation. This implies that
the number of units allocated to $i$ after the deviation has to be greater
than the number of units allocated to $j$ before the deviation. But again,
the pricing function for $i$ after the deviation is worse than the pricing
function for $j$ before the deviation. For $i$ to be allocated more, her
budget spent after the deviation has to be greater than $j$'s budget spent
before the deviation, which is a contradiction.\medskip 

Suppose $i$ is currently a loser and deviates to $v_{i}^{+}$ and becomes a
full winner. We can split this into two deviations. First, $i$ deviates to $%
v_{i}^{+\prime }>v_{j}$ and becomes a partial winner (which increases the
revenue), then she deviates to $v_{i}^{+}$ and becomes a full winner which
will next be shown to increase the revenue.\medskip 

\item Lastly, consider bidder $i$ who is a partial winner by bidding $\left(
b_{i},v_{i}\right) .$ It is obvious that she cannot become a loser after
deviating to $v_{i}^{+}.$ If she deviates to $v_{i}^{+}$ and remains a
partial winner, then the pricing function for all winners get worse, hence
the revenue has to increase. If she deviates to $v_{i}^{+}$ and becomes a
full winner, then we argue that revenue has to increase.\medskip 

Consider the case where $i$ is currently the partial winner, and she
deviates to $v_{i}^{+}>v_{i-1}$ (where bidder $i-1$ has the next highest
value after bidder $i$) so that $i-1$ is the new partial winner and $i$ is a
full winner. Denote the original unused budget of bidder $i$ by $%
s_{i}^{\prime }$ \ and after deviation, the unused budget of bidder $i-1$ by 
$s_{i-1}^{\prime }.$ It suffices to show that $s_{i}^{\prime }\geq
s_{i-1}^{\prime }.$ Assume for contradiction that $s_{i-1}^{\prime
}>s_{i}^{\prime }.$ First, it is easy to see that the pricing function for
all winners other than $i$ or $i-1$ gets worse, therefore they will be
allocated (weakly) less number of items. Similarly to the previous
discussion, we show that the total number of units allocated to bidder $i$
and $i-1$ has to (strictly) decrease after the deviation, which gives us the
desired contradiction. Bidder $i-1$'s allocation is decreased by%
\begin{equation*}
x\left( \alpha ^{c},b_{i-1}\right) -x\left( \alpha ^{c+s_{i}^{\prime
}},b_{i-1}-s_{i-1}^{\prime }\right) 
\end{equation*}%
which is strictly greater than%
\begin{equation*}
x\left( \alpha ^{c},s_{i}^{\prime }\right) .
\end{equation*}%
Bidder $i$'s allocation is increased by at most%
\begin{equation*}
x\left( \alpha ^{c+s_{i}^{\prime }-s_{i-1}^{\prime }},b_{i}\right) -x\left(
\alpha ^{c+s_{i}^{\prime }},b_{i}-s_{i}^{\prime }\right) 
\end{equation*}%
which is smaller than%
\begin{equation*}
x\left( \alpha ^{c+s_{i}^{\prime }-s_{i-1}^{\prime }},s_{i}^{\prime }\right)
.
\end{equation*}%
Since $c+s_{i}^{\prime }-s_{i-1}^{\prime }<c,$ we conclude that the total
number of units allocated to players has to be strictly less than $m$,
leading to a contradiction.
\end{itemize}
\end{itemize}

\end{document}